\documentclass[acmsmall,screen,nonacm]{acmart}
\usepackage{graphicx} 
\usepackage{algorithm}
\usepackage{algpseudocode}
\usepackage{footmisc}
\usepackage{xparse}
\usepackage{xcolor}
\usepackage{listings}
\usepackage{booktabs}
\usepackage{pgfplotstable}
\usepackage{enumitem}
\usepackage{color, colortbl}
\usepackage{float}
\usepackage{etoolbox}  
\usepackage{subcaption}
\usepackage{tcolorbox}
\usepackage{xspace}
\usepackage{makecell}
\usepackage{multirow}
\usepackage{array} 
\usepackage{makecell}

\definecolor{Cyan}{rgb}{0.88,1,1}

\definecolor{codeblue}{RGB}{0,110,179}
\definecolor{codegreen}{rgb}{0,0.6,0}
\definecolor{codered}{RGB}{220,20,60}
\definecolor{codegray}{rgb}{0.5,0.5,0.5}
\definecolor{codepurple}{rgb}{0.58,0,0.82}
\definecolor{codegold}{RGB}{106,128,0}
\definecolor{bg_green}{RGB}{252,255,245}
\definecolor{bg_blue}{RGB}{0,51,102}

\tcbuselibrary{listings}

\tcbset{
  fonttitle=\footnotesize,
  center title,
  colback=bg_green,
  colframe=bg_blue,
  listing only,
  listing options={
    basicstyle=\tt\scriptsize,
    backgroundcolor=\color{bg_green},   
    commentstyle=\color{codegreen},
    keywordstyle=\color{magenta},
    stringstyle=\color{codepurple},
    numbers=none,
    deletekeywords={struct,for,do,case},
    breakatwhitespace=false,
    breaklines=true,
    keepspaces=true,
    showspaces=false,
    showstringspaces=false,
    showtabs=false,
    tabsize=4,
    aboveskip=0pt,
    belowskip=0pt
  },
  boxrule=1pt,
  arc=2mm,
  toptitle=-0.1em,
  bottomtitle=-0.2em,
  lefttitle=0.1em,
  righttitle=0.2em,
  top=1pt,
  bottom=1pt,
  left=-0.02em,
  right=-0.2em
}

\makeatletter
\let\old@lstKV@SwitchCases\lstKV@SwitchCases
\def\lstKV@SwitchCases#1#2#3{}
\makeatother
\usepackage{lstlinebgrd}

\newcommand{\redlines}{}  
\newcommand{\greenlines}{} 

\newcommand{\setredlines}[1]{  
  \renewcommand{\redlines}{#1}  
} 
\newcommand{\setgreenlines}[1]{  
  \renewcommand{\greenlines}{#1}  
} 

\newcommand{\ifred}[1]{%
  \edef\doHighlight{0}%
  \renewcommand*{\do}[1]{%
    \ifnumequal{##1}{\value{lstnumber}}{%
      \edef\doHighlight{1}%
      \listbreak%
    }{}%
  }%
  \expandafter\docsvlist\expandafter{\redlines}%
  \ifnumequal{\doHighlight}{1}{#1}{}%
}  

\newcommand{\ifgreen}[1]{%
  \edef\doHighlight{0}%
  \renewcommand*{\do}[1]{%
    \ifnumequal{##1}{\value{lstnumber}}{%
      \edef\doHighlight{1}%
      \listbreak%
    }{}%
  }%
  \expandafter\docsvlist\expandafter{\greenlines}%
  \ifnumequal{\doHighlight}{1}{#1}{}%
}

\makeatletter
\NewDocumentCommand{\LeftComment}{s m}{%
  \Statex \IfBooleanF{#1}{\hspace*{\ALG@thistlm}}\(\triangleright\) #2}
\makeatother
\title{Enabling Memory Safety of C Programs using LLMs}

\author{Nausheen Mohammed}
\affiliation{%
  \institution{Microsoft Research}
  \country{India}
}

\author{Akash Lal}
\affiliation{%
  \institution{Microsoft Research}
  \country{India}
}

\author{Aseem Rastogi}
\affiliation{%
  \institution{Microsoft Research}
  \country{India}
}

\author{Subhajit Roy}
\affiliation{%
  \institution{Indian Institute of Technology, Kanpur}
  \country{India}
}

\author{Rahul Sharma}
\affiliation{%
  \institution{Microsoft Research}
  \country{India}
}

\pgfplotsset{compat=1.18} 

\newcommand{\tool}{\texttt{MSA}\xspace}

\newcommand{\threec}{\textbf{3C}\xspace}

\newcommand{\ptr}{\texttt{ptr}}
\newcommand{\arr}{\texttt{arr}}
\newcommand{\ntarr}{\texttt{nt\_arr}}
\lstdefinestyle{promptstyle}{
    backgroundcolor=\color{Cyan},   
    basicstyle=\ttfamily\footnotesize,
    breaklines=true,                 
    captionpos=t,
    breakindent=0pt,                    
    showtabs=false,                  
}
\definecolor{commentgreen}{RGB}{0,95,0}

\lstdefinestyle{cstyle}{
    basicstyle=\ttfamily\small,
    identifierstyle={\ttfamily},
    morekeywords=[1]{arr, nt_arr},
    keywordstyle=[1]{\ttfamily\color{purple}},
    keywordstyle=[2]{\ttfamily\color{purple}},  
    breaklines=true,                 
    captionpos=t,
    linebackgroundcolor={\ifred{\color{red!15}}\ifgreen{\color{green!15}}},
    commentstyle=\color{commentgreen}\upshape
}

\begin{document}

\begin{abstract}
Memory safety violations in low-level code, written in languages like C, continues to remain one of the major sources of software vulnerabilities. One method of removing such violations by construction is to port C code to a safe C dialect. Such dialects rely on programmer-supplied annotations to guarantee safety with minimal runtime overhead. This porting, however, is a manual process that imposes significant burden on the programmer and, hence, there has been limited adoption of this technique. 

The task of porting not only requires inferring annotations, but may also need refactoring/rewriting of the code to make it amenable to such annotations. In this paper, we use Large Language Models (LLMs) towards addressing both these concerns. We show how to harness LLM capabilities to do complex code reasoning as well as rewriting of large codebases. We also present a novel framework for \textit{whole-program transformations} that leverages lightweight static analysis to break the transformation into smaller steps that can be carried out effectively by an LLM. We implement our ideas in a tool called \tool that targets the CheckedC dialect. We evaluate \tool on several micro-benchmarks, as well as real-world code ranging up to $20$K lines of code. We showcase superior performance compared to a vanilla LLM baseline, as well as demonstrate improvement over a state-of-the-art symbolic (non-LLM) technique.  
\end{abstract}

\maketitle

\section{Introduction}
Legacy C-programs are pervasive, which makes memory corruption vulnerabilities a major problem for software systems. This problem has attracted a wealth of attention for decades but memory safety violations continue to remain one of the major sources of cyber attacks. 
From {\tt memorysafety.org}:
``Microsoft estimates that 70\% of all vulnerabilities in their products over the last decade have been memory safety issues. Google estimated that 90\% of Android vulnerabilities in the wild are memory safety issues. An analysis found that more than 80\% of the exploited vulnerabilities were memory safety issues."

Researchers have proposed safe dialects of C, such as Checked-C~\cite{checkedc}, Deputy~\cite{deputy}, Cyclone~\cite{cyclone}, etc. These all use static analysis and lightweight runtime checks to ensure formal memory safety guarantees at low runtime overheads. However, these techniques require source-level annotations. The manual cost of adding these annotations, along with the code rewriting that enable such annotations in the first place, are the main hurdle for adoption of these dialects.


Recently, Large Language Models (LLMs) have shown promise in improving the productivity of software developers \cite{10.1145/3633453}. LLMs are highly versatile and accomplish diverse tasks surprisingly well, given the right instructions as prompts. 
Motivated by their novel capabilities, we present \tool, a tool that leverages LLMs to help port C to Checked-C.  
We are unaware of any prior LLM or Machine Learning based approach for this task. 
Although recent symbolic analyses have shown promising results~\cite{3c}, our evaluation shows that they miss out on many annotations and do not perform refactoring. 

Our main contribution is a novel framework that tightly couples LLMs and symbolic representations. We show a general recipe of breaking whole program transformations  into smaller tasks that can fit into LLM prompts, where each task has a code snippet and a symbolic context that contains relevant information about the other parts of the program that are not included in the snippet (Section~\ref{sec:wpt}). \tool works in multiple phases, where each phase is an instantiation of this recipe for a different task. We present the prompts that we used for each phase (Section~\ref{sec:tech}).

LLMs help compensate for shortcomings in symbolic inference techniques by dealing with complex code patterns, as well as performing refactoring where needed to allow for further annotations in the code. For instance, if a procedure signature is modified, then the effects have to be propagated to its caller, possibly transitively. Our evaluation shows that LLMs are able to accomplish this task well when provided a sufficiently detailed prompt. On the other hand, LLMs can hallucinate, thus, using symbolic information where available helps improve overall accuracy.
Finally, note that any annotations generated by our tool are checked by the Checked C compiler. Hence, even if the LLM falters and generates a wrong annotation, safety is not compromised. The higher the accuracy of \tool, the less is the amount of time developers have to spend porting to Checked C.

Although this paper focuses on memory safety, we believe our contributions have wider ramifications in formal verification of real-world software. At a high level, any such verification task requires three steps. First step is to rewrite the code so that it is amenable to verification. Second step is to annotate the code with contracts (invariants, type qualifiers, etc.). The third and final step is to check that the code satisfies the contracts, usually with an automated verifier. These steps are interconnected, and any failure in the third step has to be fixed by repeating the first two. While symbolic techniques have been devised towards the second step, for inferring contracts, the first step has generally not received much attention. Given that LLMs have shown great promise in tedious programming tasks, it is a natural research direction to explore whether they can help with the first two steps, leaving the third untouched in order to avoid compromising soundness. Our work answers this question in the affirmative for the task of memory safety of C. 

To summarize, we make the following contributions:
\begin{itemize}
    \item We present \tool, the first LLM-based assistant for porting C to Checked-C. \tool performs transformations that are out-of-reach of existing (symbolic-only) assistants.
    \item We present a novel recipe for breaking a whole program transformation into smaller tasks that can fit into LLM prompts.
    \item We evaluate \tool on real world C-programs, ranging up to $20$K lines of code, showing that it can successfully infer $86$\% of the required annotations correctly. 
\end{itemize}

The rest of this paper is organized as follows. Section~\ref{sec:challenges} provides a background on Checked C, followed by examples that illustrate the challenges of the porting process from C code. Section~\ref{sec:threec} provides background on the state-of-the-art symbolic tool for Checked C inference. Our technical contributions follow next. We provide our generic recipe for whole program transformations using LLMs (Section~\ref{sec:wpt}) and
then we show how \tool instantiates this recipe to overcomes the challenges in the porting process (Section~\ref{sec:tech}). We evaluate \tool (Section~\ref{sec:eval}), discuss threats to validity (Section~\ref{sec:threats}), and survey related work (Section~\ref{sec:related}).

\section{Challenges in porting C to Checked C}
\label{sec:challenges}

Checked C is a safe dialect of C, inspired from Deputy \cite{deputy} and Cyclone \cite{cyclone}. It differs from its predecessors in that it allows \textit{checked} and \textit{unchecked} code to coexist. Checked regions guarantee \textit{spatial} memory safety, i.e, any illegal out of bounds memory access is caught and the program is terminated. More precisely, Checked C satisfies a \textit{blame} property where any illegal access can be blamed on the unchecked parts of the code \cite{checkedc}. 

\paragraph*{Checked Pointer Types}
Checked C introduces three checked pointer types, namely, \ptr{}$\texttt{<T>}$, \arr{}$\texttt{<T>}$ and \ntarr{}$\texttt{<T>}$ in place of the C pointer type \texttt{T*}. (These names are abbreviated from their actual names, \texttt{\_Ptr}$\texttt{<T>}$, \texttt{\_Array\_ptr}$\texttt{<T>}$ and \texttt{\_Nt\_Array\_ptr}$\texttt{<T>}$ for brevity.) The \ptr{} type is used for pointers that point to a single object (or null) and are not involved in pointer arithmetic. The compiler inserts a null check at every dereference of a \ptr{} type for spatial safety. 

The \arr{} type is used for pointers that point to an array of values. It is accompanied by a bounds expression that specifies the range of memory that the pointer can access. These declarations appear as \arr{}$\texttt{<T> p :~bounds(lo, hi)}$ where \texttt{lo} and \texttt{hi} are expressions that evaluate to the lower and upper bounds of the array, respectively. In addition to the null check, a bounds check of $\texttt{lo <= (p+i) \&\&  (p+i) < hi }$ is also inserted at every dereference \texttt{*(p+i)} for an \arr{}. There are other shorthand annotations, such as $\texttt{count(n)}$ or $\texttt{byte\_count(n)}$ that specify the number of elements or bytes that the pointer can access, starting from its current value. These checks are inserted at higher-level passes of the compiler, and may get optimized away by lower-level passes in the compiler if it manages to prove that accesses are within the supplied bounds. This combination of type-assistance and low-level optimizations makes the Checked C approach appealing compared to other safe C approaches; portions of the FreeBSD kernel that were ported to Checked C reported essentially no runtime overhead \cite{DBLP:conf/secdev/DuanYZC20}.

The \ntarr{} type is used for pointers that point to null-terminated arrays, mostly strings. An annotation of \texttt{count(n)} implies that the array has at least $\texttt{n}+1$ elements, the last being the null value. An interesting feature of \ntarr{} is that its bounds can be widened until a null character is found. Hence, \texttt{while(*p != 0) p++;} is a valid way to access \ntarr{}$\texttt{<T>}$ \texttt{p:count(0)}. Each time \texttt{*p} does not equal the null character, its bound can be widened by $1$.

\paragraph*{Challenges}
Next, we explain some of the challenges in porting C to Checked C through examples. In the changelogs below, the C-code in red needs to be replaced with code in green for a successful port. The examples in this section are derived from real world C programs used in our evaluation, and they are beyond the capabilities of existing inference tools (Section~\ref{sec:threec}).

The first challenge is in handling nested pointers.
Given a nested pointer, e.g., \lstinline{long** pt}, it is not possible in Checked C to separately annotate the buffers \lstinline{pt[0]}, \lstinline{pt[1]}, etc., with their sizes. The idiomatic way to handle this situation is to replace the nested pointer with an array of structs, where the struct has a buffer with its associated length. Correspondingly, every access to \lstinline{pt} throughout the program must be modified to respect this new interface. A real-world example follows:
\setredlines{4,8}
\setgreenlines{5,6,9,11}
\begin{lstlisting}[style=cstyle]
void AllocAssign(void) {
  ulong net; ulong n = channelNets+1;
  costMatrix = 
-   (long**) malloc(n * sizeof(long *));
+   (arr<struct arr_of_long>)
+   malloc(n * sizeof(struct arr_of_long));
  for (net = 1; net <= channelNets; net++) {
- costMatrix[net] = 
+ costMatrix[net].ptr = 
    malloc((channelTracks+2) * sizeof(long));
+ costMatrix[net].len = channelTracks+2;
  }
}
\end{lstlisting}
\setredlines{}
\setgreenlines{}
Here, \lstinline{costMatrix} has been converted from a nested pointer to an array of structs, where each struct has a buffer \lstinline{ptr} and its associated length \lstinline{len}. Whenever the buffers in \lstinline{costMatrix} are modified, the newly introduced size \lstinline{len} must be updated correctly as well. 

Next, we show an example where annotating a buffer with its bounds requires arithmetic reasoning involving a loop. Consider the following code:
\setredlines{2}
\setgreenlines{3}
\begin{lstlisting}[style=cstyle, numbers=none]
static void byteReverse(
- unsigned char* buf,
+ arr<unsigned char> buf: count(longs * 4), 
unsigned longs) {
  uint32_t t;
  do {
    t = (uint32_t)
    ((unsigned) buf[3] << 8 | buf[2]) << 16 |
    ((unsigned) buf[1] << 8 | buf[0]);
    *(uint32_t *) buf = t; buf += 4;
  } while (--longs);
}
\end{lstlisting}
\setredlines{}
\setgreenlines{}
By analyzing this loop, we can conclude that \lstinline{buf} has a size of \lstinline{count(longs * 4)}. Existing tools \cite{3c} struggle to infer bounds that have expressions with arithmetic. 
While the above loop had deterministic behavior, the following is more complicated.

\setredlines{3,5}
\setgreenlines{4,6}
\begin{lstlisting}[style=cstyle]
struct bin_to_ascii_ret
vsf_ascii_bin_to_ascii(
- const char* p_in,
+ arr<const char> p_in: count(in_len),
- char* p_out,
+ arr<char> p_out: count(in_len * 2),
 unsigned int in_len, int prev_cr) {
 ...
 while (indexx < in_len) {
   char the_char = p_in[indexx];
   if (the_char == '\n' && last_char != '\r'){
     *p_out++ = '\r';
     written++;
   }
   *p_out++ = the_char;
   indexx++;
   ...
\end{lstlisting}
\setredlines{}
\setgreenlines{}
Here, the buffer \lstinline{p_in} must be annotated with its length \lstinline{in_len} and, for memory safety, \lstinline{p_out} must have a size of (\lstinline{in_len*2}) to handle the worst case loop behavior.

Often the annotation process is not local and requires refactoring several functions. Consider the following code:
\setredlines{2}
\setgreenlines{3,4}
\begin{lstlisting}[style=cstyle]
static int countint (lua_Integer key, 
- unsigned int* nums
+ arr<unsigned int> nums: count(count_nums),
+ int count_nums
) {
    unsigned int k = arrayindex(key);
    if (k != 0) {
      nums[luaO_ceillog2(k)]++;
      ...
\end{lstlisting}
\setredlines{}
\setgreenlines{}
Just looking at the original C code of \lstinline{countint}, it is impossible to annotate the buffer \lstinline{nums} with its size. The programmer likely has some custom invariant in mind to keep the indexing within bounds. It is not possible to explain custom invariants to the Checked C compiler. Thus, the idiomatic way to enforce safety with Checked C is to introduce a new argument \lstinline{count_nums} that stores the array size and use it to annotate \lstinline{nums}, then require callers to pass the appropriate size. Let's consider a caller of \lstinline{countint} next.

\setredlines{2,4,9}
\setgreenlines{3,5,6,10}
  \begin{lstlisting}[style=cstyle]
static int numusehash (
- const Table* t,
+ ptr<Table> t,
- unsigned int* nums,
+ arr<unsigned int> nums: count(count_nums),
+ int count_nums,
  unsigned int* pna) {
  ...
- ause += countint(keyival(n),nums)
+ ause += countint(keyival(n),nums,count_nums)
  ...
\end{lstlisting}
\setredlines{}
\setgreenlines{}

Here, the call to \lstinline{countint} must be modified to match its new signature. Furthermore, \lstinline{numusehash} must annotate \lstinline{nums} with its size in its signature. Accordingly, the signature also needs an extra argument \lstinline{count_nums}. Next, the callers of \lstinline{numusehash} also need to be updated to supply this extra argument. We show one such caller next:

\setredlines{2,7,10}
\setgreenlines{3,8,11}
\begin{lstlisting}[style=cstyle]
static void rehash(lua_State* L,
- Table* t,  const TValue* ek
+ ptr<Table> t, ptr<const TValue> ek
) {
  unsigned int nums[MAXABITS + 1];
  ...
- total += numusehash(t,nums,&na); 
+ total += numusehash(t,nums,MAXABITS+1,&na);  
  ...
- na += countint(ivalue(ek),nums);
+ na += countint(ivalue(ek),nums,MAXABITS+1);
  ...
}
\end{lstlisting}
\setredlines{}
\setgreenlines{}
In the above code,  the new parameter \lstinline{count_nums} is set to its correct value \lstinline{MAXABITS+1} in the calls to \lstinline{countint} and \lstinline{numusehash}. We are unaware of any prior tool that performs such a whole program refactoring of C-code automatically.

\section{Background on symbolic inference with 3C}
\label{sec:threec}

\threec{} is a static-analysis-based tool that was proposed recently \cite{3c} to help developers port C code to Checked C. It consists of two components, a type inference algorithm called \textbf{typ3c} and a bounds inference algorithm called \textbf{boun3c}, that are executed one after the other. We briefly discuss these algorithms, their strengths and limitations since \tool uses \threec.

\textbf{typ3c} uses type qualifier inference \cite{10.1145/301618.301665} to convert legacy pointers to checked pointer types. Pointers that are used unsafely, for instance, in unsafe casts (e.g., casting an \texttt{int} to \texttt{int*}), are not converted to checked pointers. This is because the Checked C compiler cannot provide safety guarantees for such pointers. \textbf{typ3c} then classifies checked pointers into one of \ptr, \arr, and \ntarr{} depending on their use. 

Once the checked pointers have been identified, \textbf{boun3c} infers bounds for \arr{} and \ntarr{} pointers. It works by constructing a dataflow graph that tracks the flow of arrays along with their bounds, starting from their allocation site where the bounds information is available. \textbf{boun3c} is designed to be sound (i.e., inferred annotations are correct) but not complete (not all possible annotation are inferred). In fact, as our evaluation will show, \textbf{boun3c} misses several annotations. 

One limitation of \textbf{boun3c} is that it can only infer bounds that involve a single variable or a constant. Thus, bounds that involve expressions with arithmetic will be missed for sure. Further, \textbf{boun3c} is largely limited to inferring \texttt{count}-style annotations, not \texttt{bounds} annotations that have both a lower and upper bound. \textbf{boun3c} offers some heuristics that attempt to come up with more annotations, but these annotations can be unsound. Furthermore, in some cases, no good bound exists without rewriting the program, which \textbf{boun3c} is not prepared to do.  

Note that \threec, as a combination of \textbf{typ3c} followed by \textbf{boun3c}, does not attempt to fully port from C to Checked C. The resulting code, for instance, is not guaranteed to pass the Checked C compiler. This is because several bounds annotations might still be missing, as well as some dynamic casts might be necessary to pass the type checker of Checked C. \threec is, thus, an assistance in the porting process. We carry on with this design philosophy in this paper, striving to provide even more assistance without guaranteeing a complete port to Checked C. 

\section{Whole-program transformation with LLMs}
\label{sec:wpt}
As illustrated in Section~\ref{sec:challenges}, the task of porting to Checked C requires making several changes throughout the program. Even with the increasing prompt sizes, it is still unreasonable to expect entire code to fit inside a single prompt. Furthermore, we found that even when we can fit larger parts of a program in a single prompt, the accuracy of an LLM is lower when asked to make several changes, compared to doing fewer changes to a smaller piece of code (see Section~\ref{sec:eval}). 

The challenge then is to break a whole-program transformation into multiple smaller tasks. With LLMs, each inference query is treated independently of ones made previously, therefore, any query on some part of the code must provide enough context about the rest of the code in order to carry out the task effectively. We address these challenges through static analysis. 

\subsection{Dependency Graph Generation}

We use a lightweight static analysis that goes across all input source files and constructs a data structure that we call as a \textit{dependency graph}. Nodes of this graph are all the top-level declarations in the program. These can either be procedures (both its signature and its body), type declarations (\texttt{struct}, \texttt{union}, \texttt{enum}), global variable declarations or macro definitions. There is a directed edge from node $n_1$ to $n_2$ if $n_2$ is directly \textit{used} by $n_1$, as follows: 
\begin{itemize}
    \item \textbf{Procedures.} If $n_1$ is a procedure, then we place out-going edges from $n_1$ to all procedures that are directly called by it, as well as all types, globals and macros that appear somewhere inside $n_1$. We only consider direct calls. For indirect calls through a function pointer, there will be an edge to the type declaration of the pointer's type, not to potential targets of the pointer. 
    \item \textbf{Types.} If $n_1$ is a type definition, then we place out-going edges to all types and macros that appear in the definition.
    \item \textbf{Globals.} If $n_1$ is the declaration of a global variable, then we place an out-going edge to the type of the variable or any macro that the declaration might use.
    \item \textbf{Macros.} There are no out-going edges from macros. Transformation of macros is currently outside the scope of our analysis. We didn't find a need for it in our experiments.
\end{itemize}

\begin{figure}[t]
    \centering
    \includegraphics[width=0.6\textwidth]{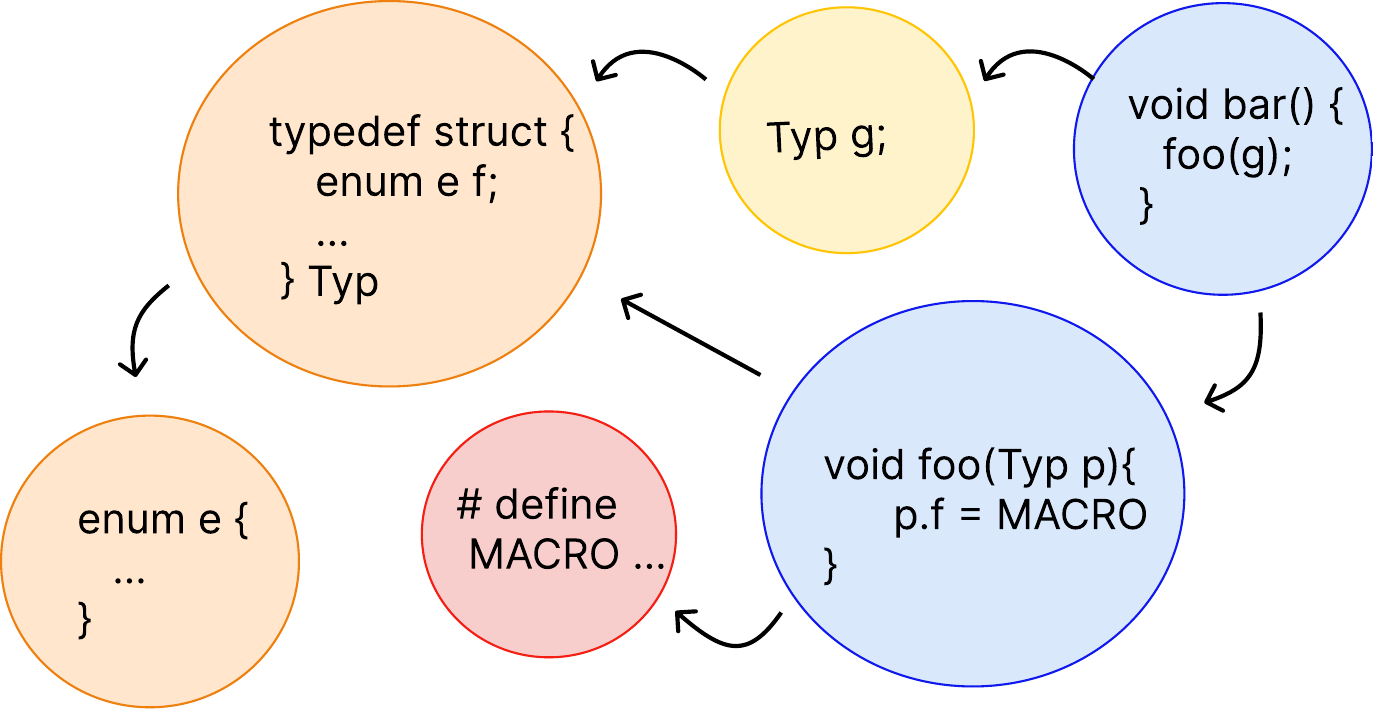}
    \caption{Example of a dependency graph}
    \label{fig:dependency-graph}
\end{figure}

The requirements of this analysis are purposefully kept simple for ease of implementation. We use \texttt{clang} to parse and construct ASTs of all input source files. We then perform linking at the AST level to resolve procedure calls and dump the dependency graph; see Figure~\ref{fig:dependency-graph} for an example.

As part of this analysis, we also record additional information for each node that can be obtained easily from its AST representation. For instance, for a procedure, we keep track of its signature, argument list, type of each argument, type of each local variable, etc. 


\subsection{Generic Whole-Program Transformation}

Porting to Checked C requires not just adding annotations, but also supporting edits that allows for the presence of an annotation in the first place. Inspired from previous experience of porting C to Checked C (like in \cite{checkedc,DBLP:conf/post/RuefLSTH19,3c}), we define three different programs transformations, which are applied sequentially in order. Each of these transformation follow a common structure, shown in Algorithm~\ref{alg:GenericTransformation}.   

\begin{algorithm}
 \caption{Whole-Program Transformation with LLMs}
 \label{alg:GenericTransformation}
\begin{algorithmic}[1]
\Procedure{ProgramTransformation}{$\mathcal{D}$, $\mathcal{T}$}
\For{$d \in \texttt{Nodes}(\mathcal{D})$}
   \State \texttt{refactored}[d] := False, \texttt{oldcode}[d] := d.code
\EndFor
\For{$d \in$ \Call{BottomUpOrdering}{$\mathcal{D}$}}
    \State prompt $\gets$ \Call{PromptTemplate}{ }
    \State prompt.task $\gets$ \Call{TaskDescription}{$\mathcal{T}$}
    \State prompt.example $\gets$ \Call{TaskExample}{$\mathcal{T}$}
    \State prompt.prelude $\gets$ d.succ
    \State prompt.code $\gets$ d.code
    \State prompt.refactor\_history $\gets$ \{(\texttt{oldcode}[u], u.code): $u$ $\in$ d.succ, \texttt{refactored}[u] \}
    \State prompt.elements $\gets$ \Call{TaskElements}{d, $\mathcal{T}$}
    \State response $\gets$ \Call{LLM}{prompt}
    \State d.code $\gets$ \Call{ApplyPatch}{response, d.code}
    \State refactored[d] $\gets$ \Call{SignatureChanged}{response}
\EndFor
\EndProcedure
\end{algorithmic}
\end{algorithm}

\begin{figure}[t]
\begin{tcblisting}{}
% (1) CheckedC Preamble
Checked C has three checked pointer types that support the following annotations:
...

% (2) Task definition
{{Task definition}}

% (3) Propagate changes
Similar changes have been made in other parts of the code. Given the refactor 
history, update the current code accordingly.

% (4) Output format
Each change must be outputted as a block with original lines and refactored
lines in the below format. Output a series of such blocks, one for every change.
...

% (5) Example
Consider this example input and output as a reference.
{{Task example}}

% (6) Code
Here is relevant context for the given code
{{prelude}}

This is the code that must be transformed
{{code}}

This is a history of the previous changes
{{refactor_history}}

Perform the given task on these parts of the code:
{{task_specific_code_elements}}
\end{tcblisting}
\caption{Prompt template for Whole-Program Transformation}
\label{fig:prompt-template}
\end{figure}

Algorithm~\ref{alg:GenericTransformation} takes the dependency graph ($\mathcal{D}$) of the input program, as well as a description of task-specific information ($\mathcal{T}$), and outputs a new program that is the result of applying $\mathcal{T}$ to the input program. The algorithm goes over the input program one declaration at a time, and instructs an LLM to preform a rewriting according to $\mathcal{T}$. 

The template of the LLM prompt is shown in Figure~\ref{fig:prompt-template}. It consists of various sections. The first section is a preamble on Checked C that defines its various annotations (e.g., \arr{}), their meaning (e.g., it is a pointer to an array) as well as the syntax rules to follow (e.g., bounds annotation appears after a colon). We share our prompts in the supplementary material.

The next section of the prompt ($2$) carries a description of the task $\mathcal{T}$ (e.g., ``infer bounds of arrays''). Next ($3$) is an instruction telling the model that prior edits have been made to the program, and it must edit the given snippet to respect those refactorings. For instance, when a callee method signature is modified, the call to that method must be modified as well.

Section $4$ of the prompt defines an output format that the model should follow. Instead of asking the model to produce the entire modified code, prior work has found it useful to instead ask for a ``diff'' or a patch that can be applied to the original code. We simply follow the prompt used in prior work \cite{rust-assistant} to obtain a patch from the LLM. Any other format or formatting instructions can be used as well. 

Section $5$ of the prompt is an example of the task. For each task, we only include one or two fixed hard-coded examples. 
Section $6$ of the prompt includes the relevant code snippets from the input program. This includes context for the current code (called ``prelude''), the code that must be transformed (code) as well as previous code refactorings (refactor\_history). Finally, the prompt also includes some code-specific elements. For instance, for the task of inferring bounds of arrays, we explicitly list the variable names with array types in order to help focus the attention of the model on those variables. 

Getting back to Algorithm~\ref{alg:GenericTransformation}, it starts (lines 2 to 4) by keeping track of the original code (\texttt{oldcode}) as well as remembering what parts of the code have been refactored (\texttt{refactored}), initially none. It then makes one pass over all declarations in the code. These declarations are picked in a bottom-up order, using a reverse topological sorting of the dependency graph. In general, the dependency graph can have cycles because of recursive types or recursive procedures; we break these cycles arbitrarily in order to limit the transformation to a single pass over the program text. 

For each declaration $d$, the prompt in instantiated with task-specific as well as code-specific details. Code prelude is computed as all immediate successors of $d$ in $\mathcal{D}$, i.e., all code elements that are referenced directly in $d$. For brevity, when including a procedure in the prelude, we only include its signature and not its body. The refactor history includes all changes made to these successors of $d$ so far, if any. 

When the LLM is prompted, it returns a patch (possibly empty) that should be applied to $d$. We apply this patch to update $d$. Finally, we set \texttt{refactored}[d] to true if the patch was non-empty, i.e, $d$ was updated. When $d$ is a procedure, we set \texttt{refactored}[d] to true only when the signature of $d$ was changed by the patch.

\begin{figure}[t]
    \centering
    \includegraphics[width=0.7\textwidth]{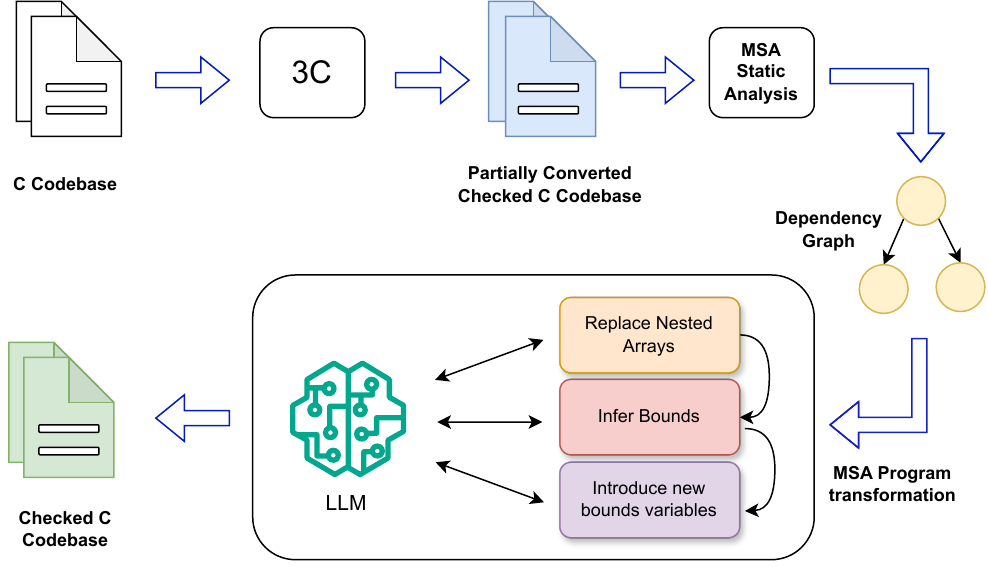}
    \caption{Workflow of porting C code to Checked C with \tool.}
    \label{fig:approach}
\end{figure}

\section{\tool Design and Implementation}
\label{sec:tech}

The design of \tool is illustrated in Figure~\ref{fig:approach}. The tool is a combination of symbolic as well as LLM-based components. 
As a design principle, we rely on symbolic components whenever they exist or are easy to implement. Tasks that are either complex to do symbolically (e.g. knowing if a code pattern respects null-termination of a given array) or require an LLM's flexibility (e.g., refactoring to accommodate a new struct field) are left to LLM-based components.

\tool feeds input C code to the \threec tool to produce partially-converted CheckedC code. \threec is quite fast (especially compared to LLM inference times), thus it serves us well to use it as a source of cheap and sound annotations. We turn off the heuristics in \textbf{boun3c}, which we observed to produce incorrect annotations. We leave heuristics to the LLM-based components of \tool. The role of \tool, thus, is to annotate the unannotated checked pointers in the program produced by \threec.

\tool takes the output of \threec, constructs its dependency graph, and uses it in subsequent LLM-based program transformations. \tool is parametric on the choice of the LLM, although we only evaluate with GPT4. In order to account for the randomness in the LLM's response, \tool asks for multiple completions (i.e., responses) for each LLM query. The default setting is $10$, although it can be changed by the user depending on their time budget. Multiple completions produce multiple code patches; \tool takes a majority vote among these patches. 

We now describe the three program transformation tasks. 

\subsection{Replacing Nested Arrays with Structs}

\begin{figure}[t]
\begin{tcblisting}{}
You are given a list of Checked C declarations and a partially converted Checked C code
snippet. Array of arr<T> is not supported in Checked C. Your task is to replace them
with an array of struct having a pointer field 'ptr' and a bounds field 'len'. You will 
also have to replace the uses of the nested array with the uses of the struct 'ptr' field
instead. Make sure to update the 'len' field whenever the 'ptr' field is updated.
\end{tcblisting}
\caption{Task description for replacing nested arrays with structs}
\label{fig:prompt-structs}
\end{figure}

\begin{figure}
\begin{lstlisting}[style=cstyle,escapeinside={(*@}{@*)}]
(*@\textbf{From:}@*)
int foo(arr<arr<int>> a, int i) {
  return a[i][i];
}
(*@\textbf{To:}@*)
// New struct
typedef struct arr_of_int {
  arr<int> ptr : count(len);
  int len;
} arr_of_int;
// type of a changes
int foo(arr<struct arr_of_int> a, int i) {
  // nested pointer access via the ptr field
  return a[i].ptr[i];
}
\end{lstlisting}
\caption{An example of the transformation (``\textbf{From}'' to ``\textbf{To}'') for nested arrays that is provided to the LLM. The code comments are included as well.}
\label{fig:ExampleNestedArrays}
\end{figure}

As mentioned in Section~\ref{sec:challenges}, the type \arr$\texttt{<\arr<T>>}$ is not allowed in Checked C syntax. The recommended way is to replace it with \arr$\texttt{<struct arr\_of\_T>}$, where \texttt{struct arr\_of\_T} is a new structure with two fields, one of type \arr$\texttt{<T>}$ for storing the inner array, and the other of type \texttt{int} for storing the bound of this array. The rest of the program should change to use this structure. For instance, the bounds field must be updated when the corresponding array field is set or updated. 


Figure~\ref{fig:prompt-structs} shows the description for this task that is used in conjunction with the template shown in Figure~\ref{fig:prompt-template} to carry out the program transformation. \tool first adds the declaration of \texttt{struct arr\_of\_T}, once for each type \texttt{T} such that the type \arr$\texttt{<\arr<T>>}$ appears in the input program. This step is done symbolically, at the AST level. \tool then uses Algorithm~\ref{alg:GenericTransformation} to carry out the required transformations in the program to use this new type. Figure~\ref{fig:ExampleNestedArrays} shows the example that is included in the prompt. The task-specific elements that are included in the prompt are the names of variables with refactored types (e.g., \texttt{a} for the example in Figure~\ref{fig:ExampleNestedArrays}).

\subsection{Inferring Bounds Annotations}

\begin{figure}[t]
\begin{tcblisting}{}
Determine and assign 'count(..)' or 'bounds(.., ..)' expressions for each att and nt_arr in
the given function. To find valid bounds for a pointer p, examine all uses of p and set
bounds that encompass every access. Alternatively, adopt the bounds from the pointer from
which p was assigned.   

You will be provided a list of pointer variable names along with their declaration line
number. You must choose one of the following rules for each of them.

[A0] Infer a valid bounds expression:
    Provide a 'count(..)' or 'bounds(..,..)' expression at the line of declaration. Choose
    this only when you are completely sure that the bounds are valid. 

[A1] Say unknown:
    When there is not enough information to infer bounds for a pointer, it is okay to leave
    the annotated line same as the original line. Follow this by explaining why enough 
    information is not available. This can be chosen when there is not a clear upper bound
    to all accesses through the pointer or the pointer depends on other pointers whose 
    bounds are not known.

[A2] Change an arr to nt_arr:
    If you cannot infer the bounds to arr p but you do know that p is terminated with a
    null character from its use, you can change its type to nt_arr. Make sure to also 
    change the pointers that p was derived from to nt_arr in such a case. This can also
    be due to a callee now taking nt_arr instead of arr due to an earlier refactor.

[A3] Add a parameter for bounds:
    If you cannot infer a reasonable bound for a pointer parameter, add a new parameter to
    store its bounds and use that in the bounds expression. Going ahead, all calls of this
    function will have to be passed this extra bounds argument.
\end{tcblisting}
\caption{Task description for inferring bounds}
\label{fig:prompt-bounds}
\end{figure}

\begin{figure}
\begin{lstlisting}[style=cstyle,escapeinside={(*@}{@*)}]
(*@\textbf{From:}@*)
struct x { int f; int g; }
int foo(arr<struct x> a, int i) {
  int j = a[i].f;
  arr<struct x> p = a;
  return a[j].f;
}
(*@\textbf{To:}@*)
// [A3] As j is read from the heap, the access
// a[j] could be anything. Moreover, j is not
// in scope at line 1. Since 'a' is a pointer
// parameter, add a bounds parameter instead
// of saying 'unknown'.
int foo(arr<struct x> a : count(count_for_a),
  int count_for_a, int i) {
  int j = a[i].f;
  // [A0] As p is assigned a, the bounds for a
  // are valid for p too.
  arr<struct x> p : count(count_for_a) = a;
  return a[j].f;
}

(*@\textbf{From:}@*)
void foo() { 
  char a[10]; nt_arr<char> p = a;
}
(*@\textbf{To:}@*)
void foo() {
  char a[10];
  // [A0] When an array is converted to nt_arr 
  // the count is the size of the array - 1.
  nt_arr<char> p : count(9) = a;
}
\end{lstlisting}
\caption{Examples of inferring bounds for \arr{} and \ntarr{}}
\label{fig:ExampleBounds}
\end{figure}

%

The second transformation does the actual inference task of adding bounds annotations for \arr{} and \ntarr{} pointers. The LLM is asked to annotate array pointers based on their usage in the code snippet that is presented to it. Because this pass traverses the code in a bottom-up fashion, when the LLM is presented the code of a certain procedure, it will also get presented with its context. This context will include callee signatures, which have already been annotated because they came before in the dependency-graph order. This allows transitivity of the inferred annotations.

The task description for this transformation is the most detailed among the three passes and is shown in Figure~\ref{fig:prompt-bounds}. It lists down four different rules ($A0$---$A3$). The first rule ($A0$) is for inferring bounds ``when the model is sure of it'' and the second rule ($A1$) provides an escape hatch for the same. The idea behind $A1$ is to reduce hallucinations when the model is not confident of inferring bounds; that is, the model should choose to leave things unannotated rather than add an incorrect one. The third rule ($A2$) rectifies inaccuracies in \textbf{typ3c} where it fails to identify that certain arrays are null-terminated. In our experience, we found that this happens because \textbf{typ3c} is sometimes unable to identify that a comparison to the null character is being used to break out of a loop through the array. In other cases, it lacks an understanding of standard string operations (from \texttt{stdlib}), which also establish that a given array is (intended to be) null-terminated. An LLM can compensate for these limitations by promoting \arr{} to \ntarr{}, helping gain information about the length of the array. Finally, the last rule ($A3$) instructs the model that if the bound of a parameter is not obvious from its use in the code, then it should add a new parameter to the procedure, and pass the obligation to the callers (which are yet to undergo transformation) to pass the appropriate bounds in the newly added parameter. Note that the prompt only says what annotations it wants, it does not say how to obtain them. The complexity of actually doing program analysis is completely left to the model.

Figure~\ref{fig:ExampleBounds} shows two examples that are provided to the model. The first example shows applications of rules $A0$ and $A3$ for \arr{}, and the latter example shows an application of rule $A0$ for \ntarr{}, where it makes a note of the off-by-one computation for bounds of null-terminated arrays. In terms of program elements that are provided in the prompt of Figure~\ref{fig:prompt-template}, we give variable names of all \arr{} and \ntarr{} typed variables that are currently unannotated.

\tool follows Algorithm~\ref{alg:GenericTransformation} with the above task description to make changes to the program, with a few minor changes. It restricts the bottom-up traversal to only the procedures in the program, not the other top-level declarations. That is, it goes bottom-up on the program call graph. Further, when looking at a particular procedure $p$, the model is also allowed to add annotations on any globals or struct fields that $p$ uses, which are anyway present in the prelude of $p$. Finally, globals and struct fields can be used in multiple procedures, each of which are only considered one-by-one. Thus, it is possible that different procedures produce conflicting annotations for globals or struct fields. In this case, \tool detects this conflict in a post-processing step and drops the corresponding annotation. These missing annotations are left for the third transformation that follows next.

\subsection{Annotating Globals and Struct Fields}

\begin{figure}[t]
\begin{tcblisting}{}
You are given a Checked C code snippet, with a history of refactors. The refactors
introduce a new variable to store the bounds of a pointer variable, which can be a 
struct field or a global variable. Update the newly introduced bounds variable with
the correct bounds whenever its corresponding pointer variable is assigned a new 
value. Make the update in the same statement as the assignment.
\end{tcblisting}
\caption{Task description for adding new bounds variables}
\label{fig:prompt-new-vars}
\end{figure}

\begin{figure}
\begin{lstlisting}[style=cstyle,escapeinside={(*@}{@*)}]
(*@\textbf{From:}@*)
void foo(arr<struct x> a, int i) {
  a[i].p = malloc(sizeof(int) * 10);
}

(*@\textbf{To:}@*)
struct x {
  int count_for_p;
  arr<int> p: count(count_for_p);
}

void foo(arr<struct x> a, int i){
  a[i].p = malloc(sizeof(int) * 10),
  a[i].count_for_p = 10;
}
\end{lstlisting}
\caption{Example of transformation for globals and struct fields}
\label{fig:ExampleGlobalsStructs}
\end{figure}


Annotations for global variables as well as struct fields have a global scope. That is, these annotations are expected to hold throughout the lifetime of the program. Consequently, it is possible that the previous pass, which only consider one procedure at a time, fails to infer a consistent bound for them. In these cases, the third pass takes over. 

\tool creates a new variable for storing the bounds of the respectively element. In particular, it creates a new int-valued global variable called \texttt{count\_for\_g} for each global variable \texttt{g} of type \arr{} that is unannotated so far. It adds the annotation \texttt{count(count\_for\_g)} to the declaration of \texttt{g}. \tool also creates a new int-valued field called \texttt{count\_for\_f} for field \texttt{f} (in the same struct) of 
type \arr{} that is unannotated so far. It adds the annotation \texttt{count(count\_for\_f)} to the field \texttt{f}. These changes are made symbolically by \tool and added to the refactor history. \tool then runs Algorithm~\ref{alg:GenericTransformation} with the task description shown in Figure~\ref{fig:prompt-new-vars}. This transformation instructs the model to update the bounds variable each time the corresponding array is assigned. Figure~\ref{fig:ExampleGlobalsStructs} shows the example that is provided in the prompt. In terms of program elements, \tool provides the names of the global variables and struct fields for whom the additional variables have been introduced.

\section{Evaluation}
\label{sec:eval}

In this section we evaluate \tool empirically. We use gpt-4-32k from Azure OpenAI Service as the LLM in these experiments. We divide our experiments into two categories: (a) benchmarking effectiveness of the different components of our algorithm and the prompt template, and (b) evaluating the effectiveness of \tool in inferring Checked C annotations in real-world codebases. We also present our experience on porting vsftpd, one of our real-world benchmarks, end-to-end.

\subsection{Benchmarking the \tool algorithm}
\label{sec:eval-alg}

We vary different components of our algorithm and compare the results to show their relative effectiveness. We consider: (a) the background about Checked C and specific inference instructions in the prompt template, and (b) modular inference with dependency-order traversal of the codebase.

For this experiment, we use a subset of the Olden~\cite{olden} and Ptrdist~\cite{ptrdist} benchmarks, with code sizes that allow the entire program to fit in a single prompt. For large codebases, modular inference is a necessity, however, we explore if modularity helps even for small programs that may fit in a single prompt. We define a trivial algorithm that uses a prompt, consisting of some instructions followed by a full C program, and queries the LLM to obtain its output. Within this algorithm, we vary the instructions part of the prompt with the following versions:

\begin{itemize}
  \item \textbf{V0}: no background on Checked C or specific inference instructions to the LLM,
  \item \textbf{V1}: background on Checked C and inference instructions, as explained in Section~\ref{sec:wpt}.
\end{itemize}


\begin{table}[]
  \centering
  \begin{tabular}{|l|c|c|c|c|c|c|}
  \hline
       & LOC & Total & \multicolumn{2}{c|}{V1 (non-modular)} & \multicolumn{2}{c|}{\tool}\\ \cline{4-7}
       &  & &  Inf & Correct & Inf & Correct  \\ \hline
       mst & 445 & 11  & 5  & 3 (27\%) & 11 & 11 (100\%) \\
       power & 628 & 7 & 1 & 1 (14\%) & 7 & 7 (100\%) \\ 
       em3d & 702 & 22 & 13 & 11 (50\%) & 22 & 17 (77\%) \\
       anagram & 664 & 13 & 7 & 4 (30\%) & 13 & 11 (84\%) \\ \hline
  \end{tabular}
  \caption{ Comparison with a non-modular version of \tool on Olden (mst, power, and em3d) and Ptrdist (anagram) benchmarks. The numbers in parentheses indicate the percentage of total annotations inferred correctly. }
  \label{tab:vanilla}
\end{table}

Version \textbf{V0}'s performance was extremely poor, with the LLM often bailing out either by citing the problem as too hard or by saying there are no pointers to be annotated. Results for \textbf{V1} and \tool are shown in Table \ref{tab:vanilla}. For each benchmark, the table shows the total number of bounds annotations required, the number of annotations inferred by the algorithm (Inf), and the number of correct annotations amongst them (Correct). For determining which of the inferred annotations are correct, we manually prepare annotated versions of these benchmarks that compile with the Checked C compiler and pass the runtime tests. Then, an inferred annotation is defined to be correct if it matches the manual annotation in these ground-truth versions.

As the table shows, while \tool correctly infers 86\% of the annotations on-average, the non-modular version \textbf{V1} infers only 35\%. There are 5 cases across these benchmarks where refactoring instructions---absent from \textbf{V1} but present in \tool---play a role. Otherwise, the only difference is modular analysis.

\textbf{Conclusion.} Without the Checked C background and inference instructions in the prompt, LLMs cannot infer annotations even for small programs. With instructions, modular analysis is more effective than fitting the entire (small) program in a prompt. Thus, it is better to focus LLM
on one procedure at a time, providing it the dependencies in the context.






\subsection{Inferring annotations in real-world codebases}
\label{sec:eval-real}
\begin{figure}
    \centering
    \includegraphics[width=0.5\textwidth]{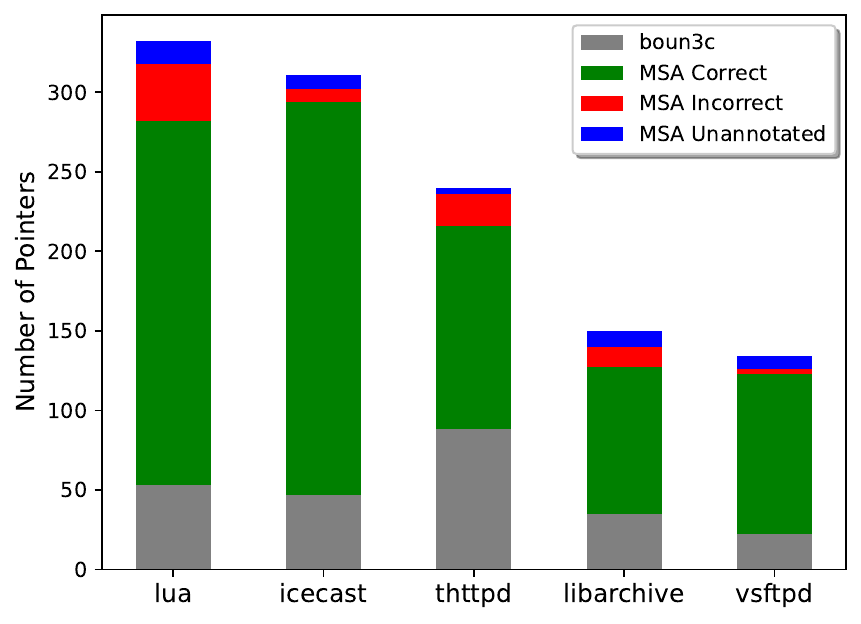}
    \caption{\tool performance on real-world codebases.}
    \label{fig:enter-label}
\end{figure}

\begin{table}[]
  \centering
  \begin{tabular}{|l|c|c|c|c|c|}
  \hline
& LOC & RB & b3c & b3ch & Remaining \\ \hline
lua & 19.4K & 332 & 53 & 31 & 279 (126 + 153) \\
icecast & 18.2K & 311 & 47 & 5 & 264 (26 + 238) \\
thttpd & 7.6K & 240 & 88 & 7 & 152 (31 + 121) \\
vsftpd & 14.7K & 134 & 22 & 15 & 112 (26 + 86) \\
libarchive & 146.8K & 150 & 35 & 3 & 115 (63 + 52)  \\ \hline
  \end{tabular}
  \caption{Benchmark details. RB is the number of required bound annotations. b3c is the number of annotations inferred by boun3c, while b3ch is the number of additional annotations inferred by boun3c heuristics. Remaining (divided into \arr{} and \ntarr{}) is the difference of RB and b3c.}
  \label{tab:sloc}
\end{table}

Table~\ref{tab:sloc} shows the details of real-world benchmarks that we picked from 3C experiments \cite{3c}, where 3C could not infer a reasonable number of annotations. For each benchmark, the table shows the size of the benchmark (in terms of lines of code), total number of pointers that required an annotation (RB), the number of pointers annotated by boun3c (b3c), additional pointers that boun3c heuristics infer (b3ch), and the number of remaining unannotated pointers (RB - b3c). We discount the boun3c heuristics from rest of our experiments; first, the number of additional annotations that they infer over boun3c is not substantial, and second, on manual inspection, nearly a third of them seemed incorrect to us. Hence, we consider Remaining as the target for \tool. For \texttt{libarchive}, we ran \tool on the whole codebase, but analyzed only a subset of the annotations for correctness (150 unannotated pointers), due to time constraints.



Figure~\ref{fig:enter-label} plots the performance of \tool on these benchmarks. We categorize the inferred annotations into Correct and Incorrect based on a manual inspection of the inferred annotations. In total, \tool infers 877 annotations across all benchmarks, out of which 797 annotations are correct and 80 are incorrect. Thus, \tool correctly infers 86\% of the annotations remaining after boun3c (797/922). Of the 797 correctly inferred annotations, 212 are array pointer bound annotations (78\% of remaining array pointer bounds) while 585 are NT array bounds annotations (90\% of remaining NT array bounds); of these 585 annotations, 532 (91\%) are \texttt{count(0)} annotations.

Below is a code snippet from \texttt{icecast} that shows a couple of annotations that \tool infers but boun3c could not:

\begin{lstlisting}[style=cstyle]
// MD5Update(...,
//   arr<char> buf:count(n), int n);
//
// util_bin_to_hex(..., int n)
//   : nt_arr<char> count (n * 2);
nt_arr<char> get_hash
  (const char *data: nt_arr<char>, int len) {
  ...
  MD5Update(&context, data, len); ...
  return (util_bin_to_hex(digest, 16));
}
\end{lstlisting}

\tool constructs the context for \texttt{get\_hash}, shown as comments, and from this information, it is able to infer that \texttt{data} has count \texttt{len} and the returned value has count \texttt{32}. boun3c is unable to infer the return type annotation because it is not able to infer the \texttt{n*2} annotation for \texttt{util\_bin\_to\_hex} (it doesn't support arithmetic expressions in annotations). It is also unable to infer annotations for \texttt{data} as one of the call sites for \texttt{get\_hash} is \texttt{get\_hash(p, strlen(p))}, and boun3c is unable to correlate the two parameters of the call.

Following is an example from \texttt{thttpd} where \tool successfully changes an \arr{} annotation from typ3c to \ntarr{}.
\begin{lstlisting}[style=cstyle]
static void defang(arr<char> str, ...) {
  arr<char> cp1 = (void *) 0;
  for(cp1 = str; *cp1 != '\0'; cp1++) {
    // access *cp1
  }
\end{lstlisting}
Here, typ3c marks the \texttt{str} argument and \texttt{cp1} as \arr\texttt{<char>}. However, analyzing the access patterns (loop iterating until the null character), \tool infers an \ntarr\texttt{<char>} type with \texttt{count(0)} annotation for both of them. There were also multiple instances when \tool refactors the code to allow additional annotations. For example:
\begin{lstlisting}[style=cstyle]
void luaL_setfuncs (arr<const luaL_Reg> l) {
  for (; l->name != NULL; l++) { ... }
} 
\end{lstlisting}
In this function, from \texttt{lua}, the array pointer \texttt{l} is incremented until its \texttt{name} field is \texttt{NULL}. Since there is no reasonable annotation for it in the code as written, \tool adds a bounds argument \texttt{l\_count} to the function and the annotation \texttt{count(l\_count)} to \texttt{l}.

Out of the 877 annotations that \tool infers, 80 (9\%) are incorrect. Most of these are subtle cases where pointers are accessed with unusual patterns. E.g., in the following snippet from \texttt{lua}, the pointer is accessed with a negative index:

\begin{lstlisting}[style=cstyle]
copy2buff(StkId top:arr<StackValue>, int n) {
  do {
    // use *(top - n)
  } while (--n > 0); 
} 
\end{lstlisting}

\tool infers \texttt{count(n)} for the \texttt{top} pointer, whereas the correct annotation would be \texttt{bound(top - n, top)}. Improving performance on such code patterns is future work.

The annotations that \tool does not infer (45/922) are mostly due to the $A1$ rule of Figure~\ref{fig:prompt-bounds}. In very few cases, the LLM produced annotations that referred to variables that are not in scope; these are dropped automatically by \tool as a post-processing step. 

\begin{table}[]
  \centering
  \begin{tabular}{|c|c|c|c|c|}
  \hline
& Functions & Globals, structs & Queries & Input, Output size \\ \hline
lua & 10 & 7 & 246 & 2627, 1895\\
icecast & 45 & 2 & 192 & 2998, 1988 \\
thttpd & 82 & 6 & 104 & 2861, 3441\\
vsftpd & 6 & 0 & 111 & 2322, 1600 \\
libarchive & 58 & 89 & 805 & 3360, 1847\\ \hline
  \end{tabular}
  \caption{Number of refactorings applied by \tool, total number of queries made to the LLM and average number of input and output tokens per query}
  \label{tab:transformations}
\end{table}

Table~\ref{tab:transformations} shows the number of function call and globals and struct refactorings that \tool applies per benchmark. The column Functions includes both the number of function parameters added as well as modifications to respective function calls. Similarly, the column for globals and structs include both the number of global variables and struct fields added as well as their corresponding assignments. For real-world codebases, the nested pointers to struct transformation did not appear (while one case of this appears in Table~\ref{tab:vanilla}).

Table~\ref{tab:transformations} also shows the number of LLM queries required per benchmark, and the average number of input and output tokens (including all completions) per query.

\textbf{Conclusion.} When porting real-world codebases to Checked C, 3C leaves a substantial number of pointers unannotated. \tool is able to infer majority of these pointers (86\% in our experiments). We observe that \tool is able to infer annotations that require sophisticated code reasoning. We also observe that porting is not just about inferring annotations; code edits, function refactorings, globals and struct refactoring are also commonplace. Their support in \tool is important for real-world codebases.





\subsection{vsftpd: End-to-end case study}
\label{sec:vsftpd}

We present a case study on end-to-end porting of \texttt{vsftpd}. This exercise takes the output of \tool and makes further edits to make the code successfully compile with Checked C. We started by reverting the incorrect annotations and adding the missing annotations in the \tool output (this includes the red and blue regions in the \texttt{vsftpd} bar in Figure~\ref{fig:enter-label}). The main work involved passing the Checked C compiler. For that, we had to change a further 148 lines in the code. These changes are the known caveats in making a codebase compile with Checked C~\cite{checkedctips}. For example, we had to add dynamic bound casts (that are checked at runtime) and assume casts, when the Checked C compiler could not reason about the bounds due to, e.g., lack of flow-sensitivity. An example is as follows:

\begin{lstlisting}[style=cstyle]
// p : nt_arr<char> count(0)
if (p[0] != '-') 
    {...} 
else 
    { // access p[1] }
\end{lstlisting}

In the \texttt{else} branch, accessing \texttt{p[1]} is safe because we know that \texttt{p} is an \ntarr{} and that \texttt{p[0]} is not null. However, the Checked C compiler is unable to reason about this, so we added an assume bound cast to bypass the checker. In a few other cases, the code was using the same pointer variable to represent different sized arrays. For such cases, we added new variables for differently sized arrays so that all variables are used in a bound-consistent manner. Once the code compiled, we were able to run the executable and start the FTP server. We were prepared to debug any runtime crashes due to failed checks inserted by Checked C (this can be caused either by a real memory-safety violation in the code, or due to incorrect dynamic casts). However, we did not encounter any such cases.  

\textbf{Conclusion.} Although this was only one case study, our experience showed that \tool annotation indeed helped. \tool roughly performed 58\% of the work: it inferred 123 annotations correctly which led to 250 edits, leaving 11 annotations for manual work, requiring 28 edits, followed by a further $148$ edits. Among these, the last set of edits were the easiest conceptually. We acknowledge that further work is required in this space to make a claim on end-to-end porting.

\section{Threats to Validity}
\label{sec:threats}

One potential threat to validity is \textit{data contamination}, if the Checked C versions of programs were part of the training data of the LLM. Given our use of LLMs from OpenAI, whose training data is not publicly known, there is no good way to completely rule it out. We note that Checked C versions of \texttt{icecast}, \texttt{thttpd} and \texttt{vsftpd} are indeed available publicly. However, there are no publicly available Checked C versions of \texttt{lua} and \texttt{libarchive}, to the best of our knowledge. The results across these benchmarks are largely consistent. Further, the LLM demonstrates poor knowledge of Checked C, unable to carry out the task without a background (\textbf{V0}, Section~\ref{sec:eval}). Our contribution of whole-program transformations in a modular fashion stands, irrespective of data contamination (\textbf{V1} vs. \tool, Section~\ref{sec:eval}). Also, Checked C code is extremely rare compared to other languages.

Another internal threat is our manual assessment of the correctness of annotations produced by \tool. We counter this by having multiple authors make independent assessment, then discussing to reach consensus. In some cases (Table~\ref{tab:vanilla} and Section~\ref{sec:vsftpd}), we validate correctness using the compiler, and found that our assessments were indeed correct.

An external threat to validity is that GPT4 can get updated over time by OpenAI, changing the performance of \tool. We make the logs of all our runs available in the supplementary material. We also make no claim on the results that will be obtained with other LLMs.

\section{Related Work}
\label{sec:related}
There are several approaches for memory safety. One set of tools use binary instrumentation to ensure safety of memory accesses~\cite{purify,valgrind,hobbes}. These do not require recompilation, but suffer from high performance overheads.
Overheads are lower for source-level instrumentation tools, at the cost of needing recompilation, but still high enough to prohibit production use~\cite{safec,ccured,DBLP:conf/pldi/NagarakatteZMZ09}. 
Cyclone~\cite{cyclone} is a safe dialect of C that extends  pointers with labels (\textit{ordinary}, \textit{never-null}, and \textit{fat}) and uses them to instrument safety checks. Deputy~\cite{deputy} provides bounded pointers and tagged unions. Checked C takes inspiration from this line of work and goes much further in building a production-ready compiler \cite{checkedcgithub}.

There has been a flurry of recent work in using LLMs to help improve the productivity of software developers. Our work builds along in this direction, while focusing specifically on memory safety through a safe C dialect. We are unaware of prior LLM-based (or ML-based) work for this problem; and we have already presented a detailed comparison against \threec, the state-of-the-art symbolic tool for this problem. 

LLMs have been extensively used for program synthesis~\cite{alphacode,codegen2022, synchromesh,sysEval22,expectation2022}, including both code generation (generating a program from a natural language description) and completion (automatically completing an incomplete code fragment). Recent auto-completion systems are even able to operate across files within a repository~\cite{liu2024repobench}. There have also been proposals that use LLMs to synthesize repair patches~\cite{Repair23,fixbugs,retrieval}, surpassing the capabilities of current automated program repair tools. 
In the direction of software security, Pearce et al.~\cite{VulnerabilityRepair2023} train their own GPT2 model to evaluate the efficacy of LLMs at generating and repairing software vulnerabilities. Their study is restricted to small ``vulnerability patterns" that lead to well-known CWE/CVEs. In the space of LLMs for software engineering, our contribution on whole-program transformations, with the ability of making several correlated edits to a program using a simple task description, is novel. We refer the reader to survey papers~\cite{survey1,survey2} for more work on leveraging LLMs for software engineering.

Recent work on a system called CodePlan \cite{DBLP:journals/corr/abs-2309-12499} 
also identifies the need to make multiple edits to a program.
CodePlan takes a set of \textit{seed} edits to a program, and then uses a custom
module to identify dependent edits that must also be made to the program, given those seed
edits. This module is based on the program AST, constructed using tree-sitter
\cite{treesitter}. In terms of comparison, CodePlan considers a generic program editing
setting. There are specific applications related to API migration and making
temporal edits, each of which are very different to the task of memory safety. 
Consequently, the algorithm used for walking program dependencies is very different between CodePlan and
\tool. Instead of completing seed edits, \tool instead focusses of inferring
annotations transitively throughout the program and doing the related program
transformations for it. 

There is also a series of work that explores the synergy of LLMs with program analysis. Llift~\cite{oopsla24} combines LLMs with static analysis tools to identify use-before-initialization bugs in the Linux kernel. Lemur~\cite{lemur} uses LLMs to guess loop invariants that are validated by a symbolic verifier. Chakroborty et al.~\cite{ranking2023} attempt to rank LLM-generated inductive invariants.  


 \section{Conclusion}
 \label{sec:concl}
 We present \tool, a tool for helping port C code to Checked C using a combination of symbolic and LLM-based techniques. \tool uses both compiler-generated ASTs as well as static analysis to extract precise knowledge about the program and help break down the complete transformation into smaller steps to effectively leverage LLMs. 
\tool uses LLMs to deal with complexity in annotating and rewriting real-world code. We also present a general algorithm for orchestrating whole-program transformations with LLMs that can be of independent interest in other domains as well. 
 Our evaluation demonstrates the ability of \tool to deal with large codebases and significantly improve performance over the state-of-the-art inference techniques for Checked C.

\bibliographystyle{ACM-Reference-Format}
\bibliography{checkedc}

\end{document}